\documentclass{aa}

\usepackage{graphicx}
\usepackage{txfonts,rotating}

\begin{document}

\title{Polarimetric survey of asteroids with the Asiago telescope
\thanks{Based on observations carried out at the Asiago Astrophysical Observatory, Italy.}}

\author{S. Fornasier \inst{1} \and
        I.N. Belskaya  \inst{2} \and
        Yu.G. Shkuratov  \inst{2} \and
        C. Pernechele \inst{3} \and
        C. Barbieri \inst{1} \and
        E. Giro \inst{4} \and
        H. Navasardyan  \inst{4}
        }

\offprints{S. Fornasier}

\institute{Dipartimento di Astronomia, Universit\`a di Padova,
Italy \email{fornasier@pd.astro.it} \and 
Astronomical Institute of Kharkiv National University, Ukraine \and INAF --
Osservatorio Astronomico di Cagliari, Italy
\and INAF -- Osservatorio Astronomico di Padova, Italy\\
}

\date{Submitted to A\&A on January 2006, Accepted for pubblication on March 2006}


\abstract{We present the first results of an asteroid
photo--polarimetry program started at Asiago--Cima Ekar
Observatory. The aim of our survey is to estimate diversity in
polarimetric properties of asteroids belonging to different
taxonomic and dynamical classes. 
The data were obtained with
the polarization analyser placed inside the Faint Object
Spectrographic Camera (AFOSC) of the 1.8m telescope. This
instrument allows simultaneous measurements of the two first
Stokes parameters without any $\lambda$/2 retarding plate. 
Our survey began in 2002, and up to now we have obtained data on a
sample of 36 asteroids; most of them are being investigated with the
polarimetric technique for the first time. Combining our data with
those already available in literature, we present an estimate of the inversion angle for 7 asteroids in this paper. Furthermore, we
present the polarimetric measurements of the rare asteroid classes 
belonging to the A and D types and a detailed VRI
observations at extremely small phase angles of the low albedo
asteroid 1021 Flammario}


\maketitle

\section{Introduction}

It is well known that asteroids are characterized by various polarimetric
properties which are similar within the same composition type (e.g. Muinonen et al. 2002).
The asteroid belt consists indeed of bodies of different
compositions, varying from primitive to strongly evolved. Primitive asteroids, such as the C--type, present
a higher value of the polarization minimum and a smaller inversion angle, as compared for instance
to S-- or E--type members.  \\
The polarization degree is defined as
P$_{r}$ = $(I_{\bot} - I_{\|})/(I_{\bot} + I_{\|})$, where $I_{\bot}$ and
$I_{\|}$ are respectively the intensity of the light scattered from the
surface perpendicular to the polarization plane and from the surface parallel to the scattering plane.  The polarization curves formed by plotting
the polarization versus the phase angle prove to be an important source of
information about the microstructure and composition of the asteroid surfaces. \\  
Particularly interesting is the so--called
negative branch of the polarization degree, in a range of phase angles between 0$^{o}$ and about 18--20$^{o}$, since it is found by laboratory measurements to be extremely dependent on the microstructure
and composition of the surface.  This negative branch has been studied for a long time by several authors and is interpreted with two main physical mechanisms, of which we report some recent investigations. \\
The first is related to the single particle scattering. It was
shown experimentally (Volten et al. 2001; Shkuratov et al.
2004) and theoretically (Shkuratov et al. 2002; Zubko et al.
2005) that irregular particles (including fractal-like
aggregates) reveal the negative polarization branch (with the
depth of 1-10\%) in a rather wide range of particle sizes and
optical constants. This polarization can survive, despite interparticle multiple scattering, even in the case of bright particulate surfaces. The reason of the negative polarization at single particle scattering has not yet been established.\\
\begin{figure}
   \centering
    \includegraphics[width=6cm]{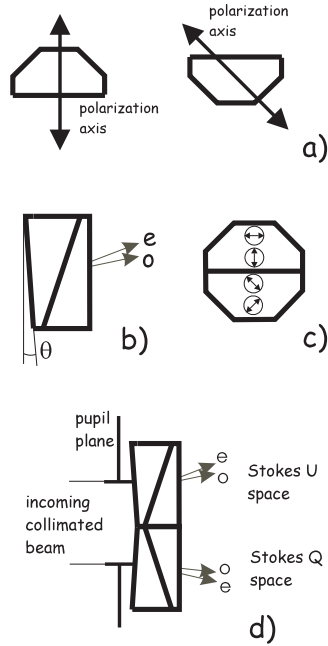}
   \caption{Views of the double Wollaston prism: a) the two separated Wollaston half-prisms; b) an edge view of one of the two Wollastons to show the wedge shape of the prism and the exit direction of ordinary and extraordinary rays; c) the prism as seen face-on, with arrows indicating the transmitted polarization components; d) the two Wollastons glued together and mounted on the re-imaged pupil.}
              \label{fig1}%
    \end{figure}
The second mechanism of the negative polarization is associated
with the coherent backscattering effect (e.g. Muinonen
2004). This effect is due to the constructive interference of
time-reversal trajectories of light multiply scattered in a
particulate surface at small phase angles. This interference can
contribute to the brightness opposition spike of
asteroids, as well as to their negative polarization branch.\\
We note that the latter mechanism is more appropriate for bright
objects, whereas the first one can provide the negative
polarization in the total albedo range. The negative
polarization strongly depends on surface albedo. Owing to that
fact, polarimetry may give independent determinations of asteroid albedos, based on empirical relations that have been well--verified in laboratory
experiments. In fact, empirical relations (Zellner \& Gradie 1976; Lupishko
\& Mohamed 1996; Cellino et al. 1999) allow for the determination
of the asteroid albedo, thanks to the knowledge of the minimum
value ($P_{min}$) of polarization or/and of the slope parameter
$h$ (i.e., the slope measured at the inversion
angle $\alpha_{0}$). Polarimetric albedos are in agreement with
data derived by other techniques, including direct measurements
such as occultations and space missions (e.g. Lupishko \&
Mohamed 1996).

Unfortunately, polarimetric observations are not easy, as polarimetric devices are offered by few telescopes. Moreover, an asteroid must be followed at different phase angles to
investigate the polarization curve in detail, and this coverage becomes difficult, considering the limited visibility of a moving target and/or weather problems in the allocated nights. 
As a consequence, although in recent years many theoretical studies and laboratory models of asteroidal polarization properties have appeared, the data set on asteroid polarimetry has not been enlarged proportionally. \\
To increase this important database, we began a survey in 2002 using a 1.8 m telescope at the Asiago Astrophysical Observatory, that was recently equipped with a polarimeter device.
In this paper, we present a description of the instrumentation together with our first results.

\section{The polarimeter of the Asiago Observatory}

\begin{figure}
   \centering
    \includegraphics[width=6cm]{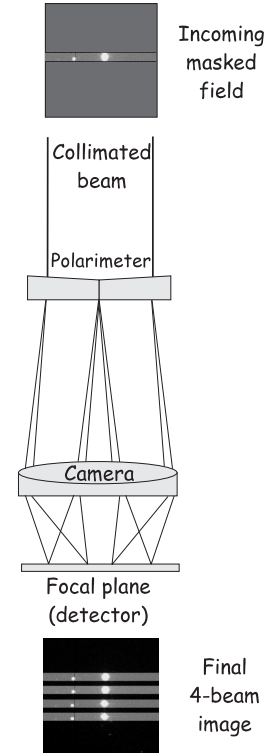}
   \caption{A schematic optical layout showing how the double prism works, the collimator is not drawn. The light coming in from the polarimetric mask on the focal plane is split in four different  beams by the polarization analyser. Finally the camera refocuses the beams on the detector.}
 \label{fig2}%
    \end{figure}

The Asiago 1.82 m telescope is equipped with the
Asiago Faint Object Spectrographic Camera (AFOSC) and a 1024x1024 pixel CCD,
providing a total field of view of 8.14x8.14 arcmin. The pixel size is 24 $\mu$m
 and the pixel scale is 0.473 arcsec/px. Since 2001, an innovative
polarimeter has been mounted on the AFOSC instrument, which allows us to simultaneously obtain images or spectra of the
astronomical source, corresponding to the information of the
polarization at $0^{o}$, $45^{o}$, $90^{o}$, and $135^{o}$,
respectively. To obtain this, two Wollaston quartz prisms have been
used, as is shown in Fig.~\ref{fig1}. \\
The Wollaston prism is made of two pieces glued together along a
plane surface perpendicular to the crystal axis of one of
them and at 45$^{o}$ to the axis of the other one (Fig. 1a,c).
The front surface of each piece is cut to obtain a wedge,
which deviates the rays above and below the optical axis (Fig. 1b). The
double prism, mounted close to the re-imaged pupil, splits the
incoming collimated beam into four beams polarized at 0, 90, 135, and 45$^{o}$, respectively (Fig. 1c,d). These four beams are
sufficient to  determine the first three elements of the Stokes
vector, i.e., the intensity I and the two linear polarization
parameters Q and U. \\
Figure~\ref{fig2} illustrates how the polarimeter works on the collimated
beam. The input field has been masked at the telescope focal plane
by a field selective mask customized for imaging or
spectropolarimetry. This is necessary to avoid the
overlap of the final images on the CCD detector. The four output beams
emerge collimated from the double Wollaston prism and are then
focused on the CCD detector by the focal camera of AFOSC.
The final image is formed by four strips from which the I, Q, and U
parameters can be extracted.
The Asiago polarimeter, can also work in the spectro-polarimetric
mode, with the Wollaston prisms mounted on the filter wheel, followed by the appropriate grism.

From the analysis of several unpolarized and polarized standard stars, it is seen that the instrumental polarization is fairly constant and always below 0.4\%, while the systematic errors in the position angle are below 1.5$^{o}$ (Desidera et al. 2004).  \\
The greater part of our observations have been obtained in the Johnson V--filter centred at 5474 \AA\ , with a bandpass of 899 \AA.

\section{Data acquisition and reduction}

 The very first results on asteroids' photo--polarimetry
were obtained in October 2002; subsequently, we had several observing runs allocated in
service mode (see Tables 1 and 2).

\begin{table*}
   \caption{Results of the polarimetric observations in the V filter. Cl is the spectral classification according to the Tholen taxonomy (Tholen 1989), while $\alpha$ and $\phi$ are, respectively, the phase and position angles.}
\begin{center}
\label{tab1}
\begin{tabular}{|l|c|c|l|l|c|c|c|c|c|} \hline
Asteroid & DATE & UT & $\phi$ ($^{o}$)  & $\alpha$ ($^{o}$) & P (\%) & $\theta$ ($^{o}$) & P$_r$ (\%) & $\theta_r$ ($^{o}$) & Cl \\ \hline
15  Eunomia & 2002-10-29 & 20:13 & 80.4 & 21.4 & 0.09$\pm$0.02 & 156.0$\pm$6.4 & +0.08$\pm$0.030 & 165.6$\pm$6.4 & S \\
21  Lutetia & 2004-12-14 & 20:58 & 68.5 & 18.8 & 0.69$\pm$0.07 & 86.1$\pm$2.9 & -0.56$\pm$0.07 &  107.7$\pm$2.9 & M \\
21  Lutetia & 2005-03-17 & 19:18 & 76.8 & 20.3 & 0.54$\pm$0.07 & 60.6$\pm$3.7 & -0.46$\pm$0.07 & 73.8$\pm$3.7 & M \\
44  Nysa    & 2002-10-30 & 01:15 & 272.4 & 10.6 & 0.25$\pm$0.08 & 100.5$\pm$9.0 & -0.24$\pm$0.08 & 98.5$\pm$9.0 & E \\
46  Hestia & 2002-10-29 & 23:48 &  268.0 & 5.1 & 1.38$\pm$0.08 & 86.3$\pm$1.7 & -1.37$\pm$0.09 & 88.3$\pm$1.7 & P \\
71  Niobe & 2003-09-27 & 00:19 & 122.6 & 8.2 & 0.39$\pm$0.09 & 113.3$\pm$6.6 & -0.37$\pm$0.09 & 80.7$\pm$6.6 & S \\
71  Niobe & 2003-12-14 & 20:05 & 65.1 & 18.3 & 0.12$\pm$0.03 & 159.1$\pm$7.1 & +0.12$\pm$0.03 & 3.9$\pm$7.1 & S \\
102  Miriam & 2003-12-15 & 00:02 & 339.6 & 2.9 & 0.72$\pm$0.03 & 160.7$\pm$1.3 & -0.72$\pm$0.03 & 91.1$\pm$1.3 & C \\
110  Lydia & 2003-11-18  & 23:34 & 65.4 & 6.3 & 0.82$\pm$0.04 & 62.6$\pm$1.4 & -0.82$\pm$0.04 & 87.3$\pm$1.4 & M\\
110  Lydia & 2003-12-14 & 21:39 &  69.9 & 15.6 & 0.56$\pm$0.04 & 70.6$\pm$2.0 & -0.56$\pm$0.04 &  90.7$\pm$2.0 & M\\
140  Siwa & 2003-02-12 & 23:32 & 163.2 & 1.0 & 0.51$\pm$0.06 & 148.1$\pm$3.3 & -0.44$\pm$0.07 & 74.9$\pm$3.3 & P \\
140  Siwa & 2005-10-28 & 00:50 & 25.1 & 2.7 & 0.57$\pm$0.06 & 35.8$\pm$3.0 & -0.53$\pm$0.06 & 100.7$\pm$3.0 & P \\
152  Atala & 2003-11-19 & 00:12 &  77.2 & 5.3 & 0.37$\pm$0.04 & 70.4$\pm$3.9 & -0.37$\pm$0.04 & 87.0$\pm$3.9 & D/S  \\
152  Atala & 2003-12-14 & 22:50 &  73.4 & 13.7 & 0.14$\pm$0.05 & 67.3$\pm$10.0 & -0.14$\pm$0.07 & 83.9$\pm$10.0 & S \\
153 Hilda & 2003-12-15 & 03:34 & 293.9 & 5.3 & 0.90$\pm$0.10 & 108.3$\pm$3.2 & -0.88$\pm$0.10 & 84.4$\pm$3.2 & X \\
165  Lorely & 2004-12-14 & 18:35 & 69.9 & 17.6 & 0.53$\pm$0.08 & 109.8$\pm$4.3 & -0.09$\pm$0.08 &  129.9$\pm$4.3 & C/D\\
171 Ophelia & 2003-11-19 & 01:01 & 57.2 & 3.04 & 0.63$\pm$0.03 & 57.8$\pm$1.4 & -0.63$\pm$0.04 & 90.6$\pm$1.4 & C \\
208 Lacrimosa & 2002-10-29 &23:15 &  101.0 & 0.9 & 0.31$\pm$0.09 & 105.7$\pm$8.3 & -0.30$\pm$0.010 & 94.7$\pm$8.3 & S \\
214  Aschera & 2004-12-14 & 17:26 & 67.2 & 21.7 & 0.46$\pm$0.05 & 133.0$\pm$3.1 & +0.31$\pm$0.05 & 155.8$\pm$3.1 & E \\
225  Henrietta & 2004-12-15 & 03:27 & 298.0 & 12.8 & 1.30$\pm$0.14 & 122.0$\pm$3.1 & -1.29$\pm$0.14 & 94.0$\pm$3.1 & F \\
250 Bettina & 2003-11-19 & 02:09 & 106.7 & 4.8 & 0.78$\pm$0.06 & 104.9$\pm$2.2 & -0.78$\pm$0.06 & 88.2$\pm$2.2 & M\\
365 Corduba & 2003-09-27 & 01:51 & 86.8 & 2.33 & 0.84$\pm$0.03 & 88.1$\pm$1.1 & -0.84$\pm$0.03 & 91.3$\pm$1.1 & C \\
377 Campania & 2003-09-26 & 23:29 & 81.8 & 11.5 & 1.76$\pm$0.06 & 76.1$\pm$1.0 & -1.72$\pm$0.06 & 84.3$\pm$1.0 & C \\
377 Campania & 2003-12-13 & 18:03 & 67.30 & 22.5 & 0.56$\pm$0.05 & 159.3$\pm$2.5 & +0.55$\pm$0.05 & 1.9$\pm$2.5 & C \\
381 Myrrha & 2003-01-30 & 23:47 & 177.0 & 0.5 & 0.44$\pm$0.04 & 157.9$\pm$2.6 & -0.34$\pm$0.04 & 70.9$\pm$2.6 & C \\
386 Siegena & 2004-05-27 & 00:56 & 228.1 & 12.2 & 1.47$\pm$0.07 & 46.8$\pm$1.4 & -1.46$\pm$0.07 & 88.8$\pm$1.4 & C \\
386 Siegena & 2004-07-10 & 01:21 & 141.4 & 11.1 & 1.53$\pm$0.05 & 144.8$\pm$1.0 & -1.52$\pm$0.05 & 93.4$\pm$1.0 & C \\
409 Aspasia & 2002-10-29 & 22:45 & 95.1 & 7.7 & 1.56$\pm$0.05 & 95.2$\pm$1.0 & -1.56$\pm$0.05 & 90.1$\pm$1.0  & CX \\
409 Aspasia & 2003-12-15 & 05:15 & 296.3 & 14.4 & 1.04$\pm$0.07 & 115.1$\pm$2.0 & -1.04$\pm$0.07 & 88.8$\pm$2.0 & CX\\
420 Bertholda & 2004-09-18 & 23:14 & 237.3 & 8.7 & 1.03$\pm$0.05 & 58.5$\pm$1.4 & -1.03$\pm$0.05 & 91.2$\pm$1.4 & P \\
438 Zeuxo & 2004-09-19 & 01:29 & 256.6 & 15.2 & 0.83$\pm$0.05 & 64.6$\pm$1.7 & -0.76$\pm$0.05 & 78.1$\pm$1.7 & F \\
456 Abnoba & 2003-09-26 & 21:18 & 100.1 & 11.2 & 0.64$\pm$0.08 & 95.1$\pm$3.6 & -0.63$\pm$0.08 & 85.2$\pm$3.6 & S \\
456 Abnoba & 2004-12-14 & 23:06 & 70.0 & 10.1 & 0.66$\pm$0.08 & 80.5$\pm$3.5 &  -0.61$\pm$0.08 & 100.5$\pm$3.5 & S \\
466 Tisiphone & 2003-09-26 & 22:09 & 111.1 & 7.9 &   1.19$\pm$0.10 & 104.8$\pm$2.4 & -1.16$\pm$0.10 & 83.7$\pm$2.4 & C\\
466 Tisiphone & 2003-12-14 & 18:54 & 64.4 & 15.5 & 0.96$\pm$0.12 & 59.5$\pm$3.6 & -0.95$\pm$0.12 & 85.1$\pm$3.6 & C\\
466 Tisiphone & 2004-12-14 & 22:19 & 91.1 & 11.6 & 1.46$\pm$0.08 & 96.5$\pm$1.6 & -1.44$\pm$0.08 & 95.4$\pm$1.6 & C\\
466 Tisiphone & 2005-10-28 & 02:30 & 282.1 & 17.6 & 0.77$\pm$0.09 & 93.6$\pm$3.3 & -0.74$\pm$0.09 & 81.5$\pm$3.3 & C\\
762  Pulcova & 2003-11-18 & 22:34 & 77.6 & 14.1 & 0.79$\pm$0.08 & 77.8$\pm$2.9 & -0.79$\pm$0.08 &  90.1$\pm$2.9 & F\\
762  Pulcova & 2004-12-15 & 00:26 & 232.1 & 5.6 & 0.77$\pm$0.05 & 66.3$\pm$1.9 & -0.68$\pm$0.05 & 104.2$\pm$1.9 & F\\
785  Zwetana & 2003-11-19 & 03:14 & 273.1 & 4.23 & 0.81$\pm$0.04 & 95.2$\pm$1.5 & -0.81$\pm$0.04 & 92.1$\pm$1.5 & M\\
785  Zwetana & 2005-03-18 & 01:34 & 266.5 & 18.7 & 0.98$\pm$0.05 & 75.6$\pm$1.5 & -0.91$\pm$0.050 &  79.2$\pm$1.5 & M \\
863  Benkoela & 2003-05-25 & 01:33 & 151.5 & 13.7 & 0.28$\pm$0.07 & 158.2$\pm$7.0 & -0.27$\pm$0.07 & 96.7$\pm$7.0 & A \\
863  Benkoela & 2004-07-09 & 21:45 & 116.4 & 18.8 & 0.08$\pm$0.04 & 54.7$\pm$14.0 & +0.04$\pm$0.04 & 28.3$\pm$14.0 & A \\
893  Leopoldina & 2004-07-09 & 23:55 & 181.0 & 5.4 & 1.06$\pm$0.06 & 176.5$\pm$1.6 & -1.05$\pm$0.06 & 85.5$\pm$1.6 & XF \\
925  Alphonsina & 2002-10-29 & 18:10 & 78.6 & 18.5 & 0.07$\pm$0.04 & 86.0$\pm$16.0 & -0.07$\pm$0.04 & 98.2$\pm$16.0 & S \\
925  Alphonsina & 2003-12-15 & 01:37 & 173.0 & 7.9 & 0.76$\pm$0.04 & 171.2$\pm$1.5 & -0.76$\pm$0.04 & 88.2$\pm$1.5 & S \\
944  Hidalgo & 2004-12-14 & 19:30 & 80.5 & 26.8 & 2.52$\pm$0.04 & 165.8$\pm$0.5 & +2.49$\pm$0.04 &  175.4$\pm$0.5 & D\\
1021  Flammario & 2004-12-15 & 02:44 & 288 & 16.4 & 0.37$\pm$0.13 & 109.4$\pm$10.0 & -0.37$\pm$0.13 & 91.4$\pm$10.0 & F \\
1021  Flammario  & 2005-03-09 & 21:21 & 101.6 & 20.2 & 1.06$\pm$0.06 & 4.4$\pm$1.6 & +1.02$\pm$0.06 & 172.8$\pm$1.6 & F \\
1251  Hedera & 2004-12-15 & 05:15 & 290.3 & 17.0 & 0.21$\pm$0.12 & 124.5$\pm$15.0 & -0.19$\pm$0.12 & 104.2$\pm$15.0 & E\\
1627  Ivar  & 2005-03-14 & 23:23 & 135.4 & 8.7  & 0.47$\pm$0.08 & 135.4$\pm$4.8 & -0.47$\pm$0.08 & 90.0$\pm$4.8 & S\\
1627  Ivar  & 2005-03-15 & 23:35 & 133.9 & 9.2  & 0.47$\pm$0.10 & 116.0$\pm$6.0 & -0.38$\pm$0.10 & 72.1$\pm$6.0 & S\\
1627  Ivar  & 2005-03-17 & 20:45 & 131.5 & 10.3 & 0.32$\pm$0.05 & 148.0$\pm$4.5 & -0.26$\pm$0.04 & 106.5$\pm$4.5 & S \\
2797  Teucer & 2004-07-09 & 20:28 & 110.0 & 9.9 & 1.59$\pm$0.10 & 126.3$\pm$1.8 & -1.34$\pm$0.10 & 106.3$\pm$1.8 & D \\
3200  Phaeton & 2004-12-14 & 21:27 & 92.7 & 23.0 & 0.54$\pm$0.10 & 163.2$\pm$5.3 & 0.43$\pm$0.10 & 161.2$\pm$5.3 & F\\
\hline
      \end{tabular}
       \end{center}
       \end{table*}

The observational procedure includes the acquisition of
several flat field images in two different sets corresponding
to the adapter position of 0$^{o}$ and 90$^{o}$, to average
the polarization induced by the reflections on the dome screen of
the flat field lamp. Subtracting the bias, this average produces a
master flat.
Moreover, polarized and unpolarized standard stars are repeatedly
acquired during each night to calibrate the instrumental polarization
and the zero point of the position angles.
The asteroids and standard star images are then corrected for bias and master flat, and the cosmic rays removed.
The data reduction is performed by using dedicated software written in IDL. \\
The centre of the asteroid image in each of the four channels is
evaluated by a
2 dimensional centroid algorithm.  The flux for each channel is
integrated over
a radius corresponding to 3-4 times the average seeing, and
the sky is subtracted
using a 3-5 pixel wide annulus around the asteroid.  Naming the 
flux corresponding to the channels at 0, 90, 45, and
135$^{o}$ of polarization I$_{0}$, I$_{90}$,
I$_{45}$, and I$_{135}$, the Stokes parameters are then
derived in the following
manner:
\begin{equation}
  Q = \frac{I_{90}-I_{0}}{I_{90}+I_{0}}
\end{equation}

\begin{equation}
U = \frac{I_{135}-I_{45}}{I_{135}+I_{45}}
\end{equation}

The degree of polarization $P$ and the position angle $\theta$ of the polarization plane
in the instrumental reference system are expressed via the parameters Q and U with the well-known formulae

\begin{equation}
 P = \sqrt{U^2+Q^2}
\end{equation}

\begin{equation}
 \theta = \frac{1}{2} arctan\frac{U}{Q}
\end{equation}

\noindent
The errors are evaluated in the following manner: \\
\[ \sigma_{P} = \frac{|U*dU+Q*dQ|}{P} \] and  \[ \sigma_{\theta} = \frac{28.65*\sigma_{P}}{P} , \]
where dU and dQ are the errors on the Stokes parameters (Shakhovskoy \& Efimov 1972). \\
The position angle of the polarization plane relative to the
plane perpendicular to the scattering plane, $\theta_r$, is expressed in terms of the position angle  $\phi$ as
\[\theta_r = \theta -  (\phi \pm 90^{o}) , \]
 where the sign inside the bracket is chosen to assure the condition $0^o \le
(\phi  \pm 90^{o}) \le 180^o$. The polarization quantity $P_r$
is derived from \[P_r = P * cos(2\theta_r) \]
Each asteroid has been observed consecutively at least 3-5
times with an exposure time long enough to reach a high
signal--to--noise ratio ($>$ 200) in photon counts. The values reported in
Tables 1 and 2 are the means of the results of
these multiple exposures.

\begin{table*}
       \begin{center}
\label{tab3}
       \caption{Observations at small phase angles of asteroid 1021 Flammario in the V, R, and I filters.}
\begin{tabular}{|l|c|c|c|c|c|c|c|} \hline
 DATE & UT & $\phi$ ($^{o}$) & $\alpha$ ($^{o}$) & P (\%) & $\theta$ ($^{o}$) & P$_r$ (\%) & $\theta_r$ ($^{o}$) \\ \hline
V filter & & & & & & & \\ \hline
2005-01-14 & 23:12 & 284.5 & 0.6 & 0.12$\pm$0.04 & 50.0$\pm$10.0   & +0.039$\pm$0.060 & 35.5$\pm$10.0 \\
2005-01-15 & 03:08 & 284.0 & 0.5 & 0.14$\pm$0.05 & 53.5$\pm$6.0   & +0.027$\pm$0.05  & 39.5$\pm$6.0 \\
2005-01-15 & 21:00 & 272.2 & 0.1 & 0.10$\pm$0.05 & 22.8$\pm$5.0   & +0.075$\pm$0.05  & 20.6$\pm$5.0 \\
2005-01-16 & 20:39 & 109.9 & 0.4 & 0.23$\pm$0.05 & 38.2$\pm$5.0   & +0.178$\pm$0.05  & 18.3$\pm$5.0 \\
2005-01-17 & 02:15 & 109.1 & 0.6 & 0.10$\pm$0.05 & 33.3$\pm$5.0   & +0.053$\pm$0.05  & 14.2$\pm$5.0 \\
2005-01-17 & 22:30 & 108.0 & 1.0 & 0.17$\pm$0.05 & 63.4$\pm$5.0   & -0.003$\pm$0.05  & 45.4$\pm$5.0 \\ \hline
R filter & & & & & & & \\ \hline
2005-01-14 & 23:22 & 284.5 & 0.6 & 0.129$\pm$0.06 & 39.9$\pm$10.0  & +0.081$\pm$0.08 & 25.4$\pm$10.0 \\
2005-01-15 & 03:21 & 284.0 & 0.5 & 0.101$\pm$0.05 & 36.2$\pm$7.0  & +0.072$\pm$0.05 & 22.3$\pm$7.0 \\
2005-01-15 & 21:58 & 272.2 & 0.1 & 0.258$\pm$0.05 & 30.0$\pm$5.0  & +0.146$\pm$0.05 & 27.8$\pm$5.0\\
2005-01-16 & 20:52 & 109.9 & 0.4 & 0.111$\pm$0.05 & 157.6$\pm$5.0 & +0.010$\pm$0.05 & 137.6$\pm$5.0 \\
2005-01-17 & 02:27 & 109.1 & 0.6 & 0.114$\pm$0.06 & 22.9$\pm$5.0  & +0.113$\pm$0.06 & 3.8$\pm$5.0 \\
2005-01-17 & 22:37 & 108.0 & 1.0 & 0.089$\pm$0.05 & 65.7$\pm$5.0 &  -0.008$\pm$0.05 & 47.7$\pm$5.0 \\ \hline
I filter & & & & & & & \\ \hline
2005-01-14 & 23:33 & 284.5 & 0.6& 0.137$\pm$0.05 & 51.6$\pm$10.0  & +0.038$\pm$0.07 & 37.1$\pm$10.0  \\
2005-01-15 & 03:33 & 284.0 & 0.5& 0.258$\pm$0.09 & 39.4$\pm$5.0   & +0.163$\pm$0.11 & 25.4.0$\pm$5.0   \\
2005-01-16 & 21:01 & 109.9 & 0.4& 0.147$\pm$0.08 & 26.5$\pm$5.0   & +0.143$\pm$0.09 & 6.7$\pm$5.0 \\
2005-01-17 & 22:46 & 108.0 & 1.0& 0.118$\pm$0.06 & 67.2$\pm$5.0   & -0.017$\pm$0.06 & 49.2$\pm$5.0  \\
\hline
       \end{tabular}
       \end{center}
       \end{table*}

\begin{table*}
       \begin{center}
\label{tab4}
       \caption{Estimated polarimetric parameters for the asteroids which have been observed at different phase angles from this survey and literature data (IRAS albedos are taken from Tedesco et al. 2002).}
\begin{tabular}{|l|c|c|c|c|c|c|c|} \hline
Asteroid &  TYPE    & IRAS albedo & $|P_{min}|$ &  $\alpha_{inv}$  ($^{o}$) &    N & Slope (\%/deg) & Polarimetric albedo \\ \hline
   15 Eunomia   & S   & 0.21    & --        &20.6      &4   &0.087  &0.19\\
   21 Lutetia   & M   & 0.22    & --        &24.2      &8   & 0.157 &0.11\\
   71 Niobe & S, Xe   & 0.31    & --        & $<$18    &3   & 0.061 &0.27\\
 377  Campania  & PD      & 0.059   & 1.76    & 19.8     &2 &0.206  &0.08\\
 409  Aspasia   & CX      & 0.061       & $>$1.56&  19.9       &4   &0.191  &0.09\\
 466  Tisiphone & C   & 0.063   &1.6       &   --      &3   & --    &0.07\\
 863  Benkoela  & A   & 0.595   &0.4       &  18.1     &7   &0.052  &0.32\\
 925  Alphonsina& S   & 0.28    & --     &  19.6       &2   &0.065  &0.26\\

\hline
       \end{tabular}
       \end{center}
       \end{table*}

\section{Results}

We obtained data on 36 asteroids during 21 nights in 2002-2005 (Tables 1 and 2), with the aims of estimating diversity in polarimetric properties, and thus in surface characteristics. The choice of targets was made to cover asteroids of different composition and dynamical properties, which were not previously observed by the polarimetric technique. \\
Although most targets could be observed at a single phase angle, when comparing the data obtained with the available polarimetric measurements of asteroids of the same taxonomic type, we can draw interesting conclusions on the diversity of their surface properties. \\
Our approach is based on previous analyses of asteroidal polarization phase curves, which showed similar behaviour for asteroids of the
 same taxonomic class (e.g. Penttila et al. 2005). The available
database on asteroid polarimetric observations (http://pdssbn.astro.umd.edu/sbnhtml/asteroids), which contains data on about 100 asteroids of known
 taxonomic classes, has been used to supplement our observations and derive the polarimetric parameters.  These parameters are given in Table 3, which contains asteroid type, IRAS albedo,
 P$_{min}$, inversion angle $\alpha_{inv}$, number of points used to estimate polarimetric parameters,
  polarimetric slope, and polarimetric albedo defined using the {\it slope-albedo}
  or {\it Pmin-albedo} correlations adopted by Zellner \& Gradie (1976).

\begin{figure}
   \centering
    \includegraphics[width=10cm]{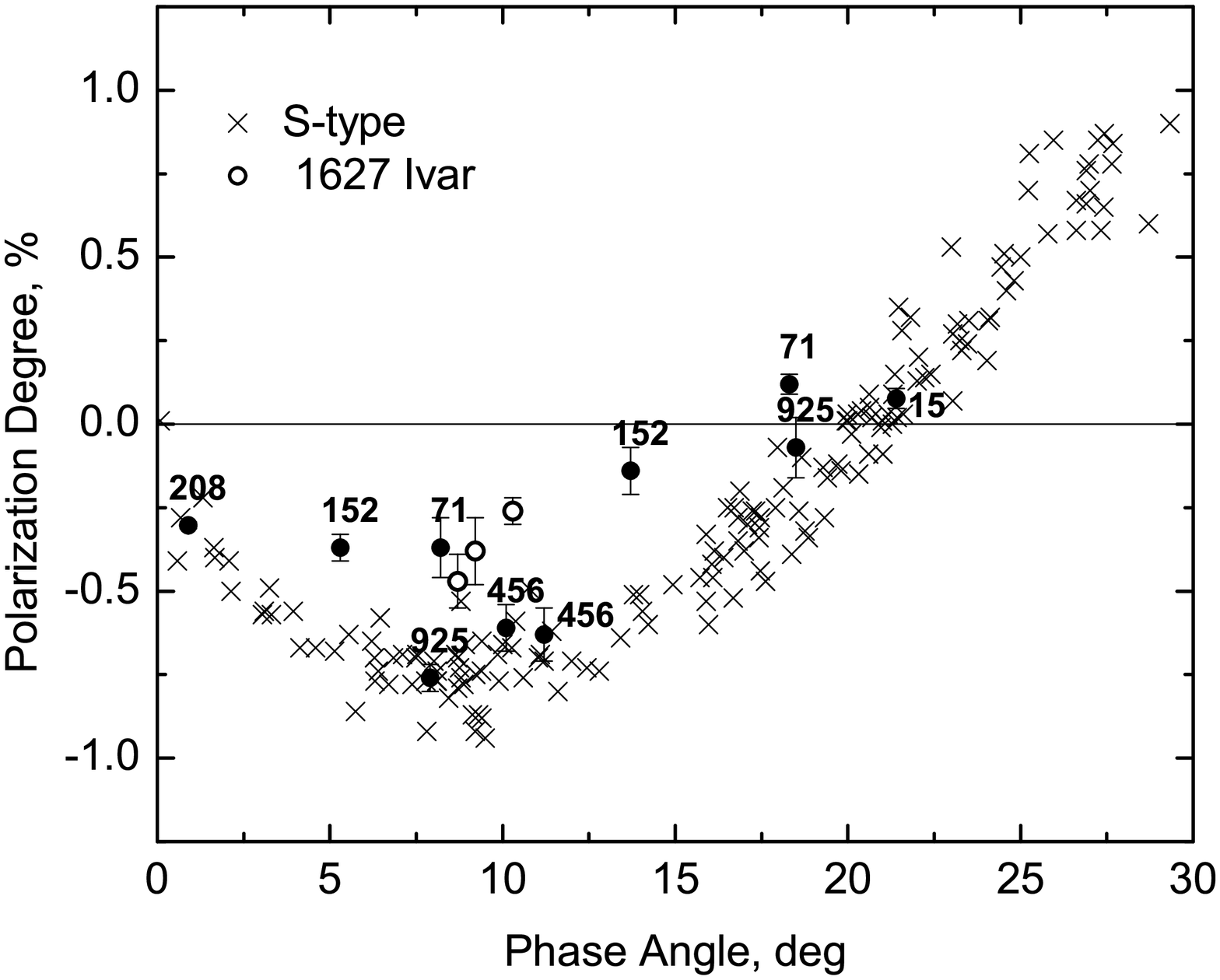}
   \caption{Polarization phase curve for the S-type asteroids. Data marked by crosses are taken from Zellner \& Gradie (1976), Belskaya et al. (2003), Broglia \& Manara (1990, 1992), and Broglia et al. (1994).}
 \label{fig3}%
    \end{figure}

    \subsection{S-type asteroids}
Seven S-type asteroids have been observed. Asteroids 15 Eunomia, 208 Lacrimosa, 456 Abnoba, and 925 Alphonsina
have polarization values within the range typical for the S-type, while asteroids 71 Niobe,
152 Atala, and 1627 Ivar are characterized by substantially low polarization degrees (Fig. 3).
Low polarization degrees imply high albedos of the surfaces of the above-mentioned asteroids. Indeed, the IRAS albedo of 71 Niobe is 0.3, and its spectral properties are close to the E-type (Xe class according to
Bus \& Binzel 2002). The albedo of 152 Atala is unknown, and there is a discrepancy in its classification:
D (Tholen 1989) and S (Bus \& Binzel 2002). Our polarimetric observations are in favor of a higher albedo than is typical for the S class. \\
Regarding the near-Earth asteroid 1627 Ivar, the polarization degree
measured near the minimum of the negative polarization branch
corresponds to an albedo of about 0.3, which is larger than that derived from the polarimetric slope (Kiselev et al. 1994) and thermal observations
(Delbo et al. 2003). Ivar becomes the third S-type near-Earth asteroid for which negative polarization has been measured. Two other asteroids, 433 Eros and 1025 Ganymed, have larger negative polarization values
typical of the S-type main belt asteroids.

\subsection{M-type asteroids}

Polarization degrees typical for the M-type asteroids have been measured for 110 Lydia and 250 Bettina
(Fig. 4).  Asteroid 785 Zwetana is characterized by a noticeable negative polarization at the phase angle of
18.7$^{o}$, implying a large inversion angle.
Its polarization is close to that of 21 Lutetia, an asteroid classified as M-type on the basis of its high IRAS albedo. However, considering the polarimetric properties (Belskaya \& Lagerkvist 1996), the lower radar albedo compared to the IRAS one (Magri et al. 1999), the 3 $\mu$m  absorption
feature diagnostic of water of hydration (Rivkin et al. 2000), and a spectral behaviour in the near
infrared region similar to that of the carbonaceous
chondrites (Birlan et al. 2004; Barucci et al. 2005), 21 Lutetia probably does not belong to the M class.
The taxonomic classification is also controversial for 785 Zwetana, due to its peculiar large inversion angle compared to that of the typical M-type asteroids. Further observations covering different phase angle aspects
are needed to fully understand the physical properties of this asteroid.
\begin{figure}
   \centering
    \includegraphics[width=10cm]{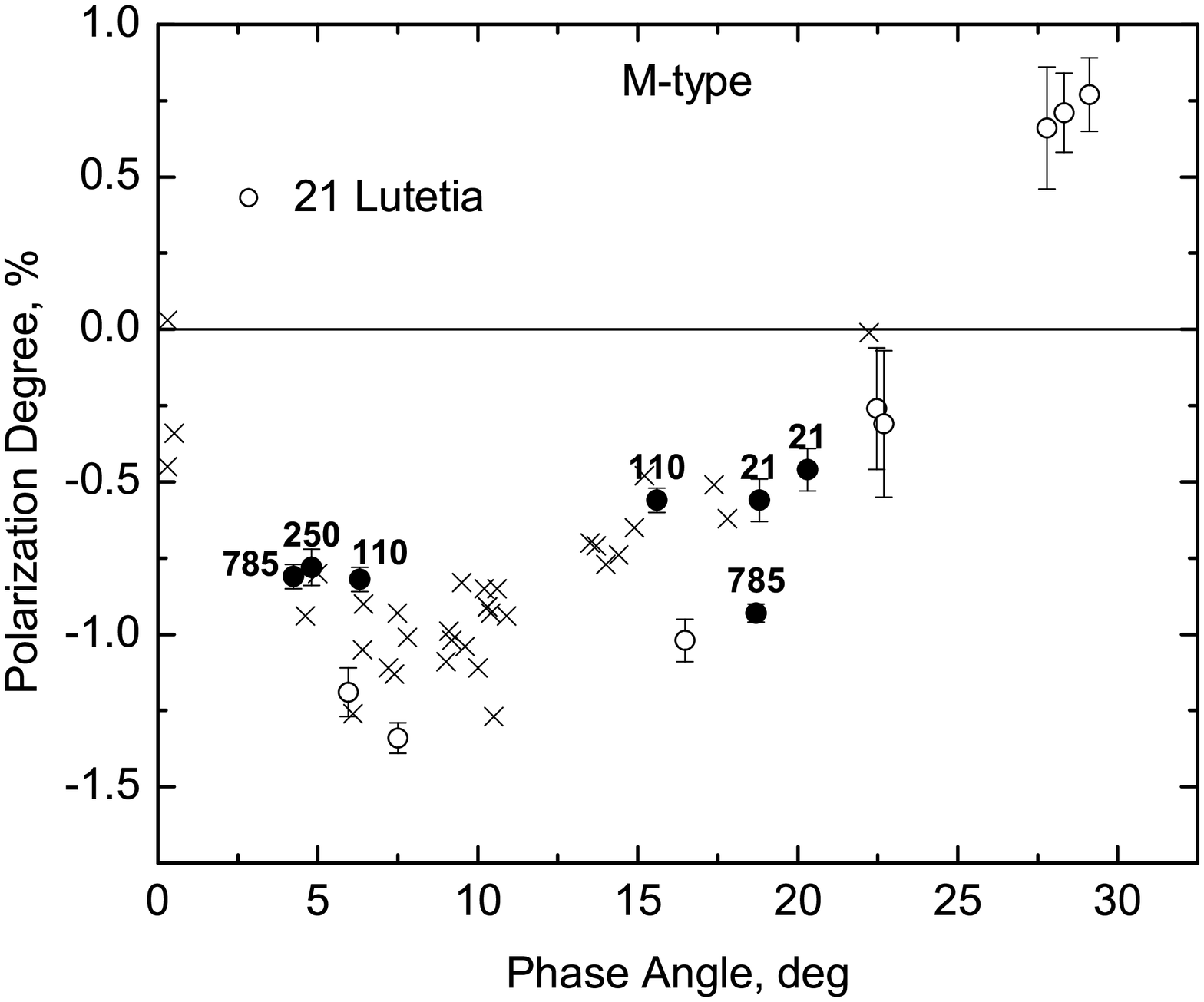}
   \caption{Polarization phase curve for the M-type asteroids. Data marked by crosses and
open circles are taken from Belskaya et al. (1985, 1987a, 1991), Broglia \& Manara (1992), Lupishko et al. (1994), and Zellner \& Gradie (1976).}
 \label{fig4}%
    \end{figure}

\subsection{E- and A-type asteroids}

We obtained polarimetric data for the E-type asteroids 214 Aschera and 1251 Hedera at a single phase angle
(at 21.7$^{o}$ and 17.0$^{o}$, respectively), close to the inversion angle of polarization.
The measured polarization values for both asteroids are typical for the E class (e.g. Zellner \& Gradie 1976; Fornasier et al. 2006) and seem to
indicate that the inversion angle is in the region of 18-19$^{o}$.\\

We have obtained the first polarimetric observations of an A-type
asteroid. We observed 863 Benkoela as did Cellino et al. (2005), independently. The observations supplement each other (see
Fig. 5) and allow us to determine the inversion angle and polarimetric
slope. The observed polarization-phase behaviour of Benkoela is
similar to that of the E-type asteroids. The estimated
polarimetric albedo is 0.32 $\pm$0.12 lower than its IRAS
albedo (see Table 3).
\begin{figure}
   \centering
    \includegraphics[width=10cm]{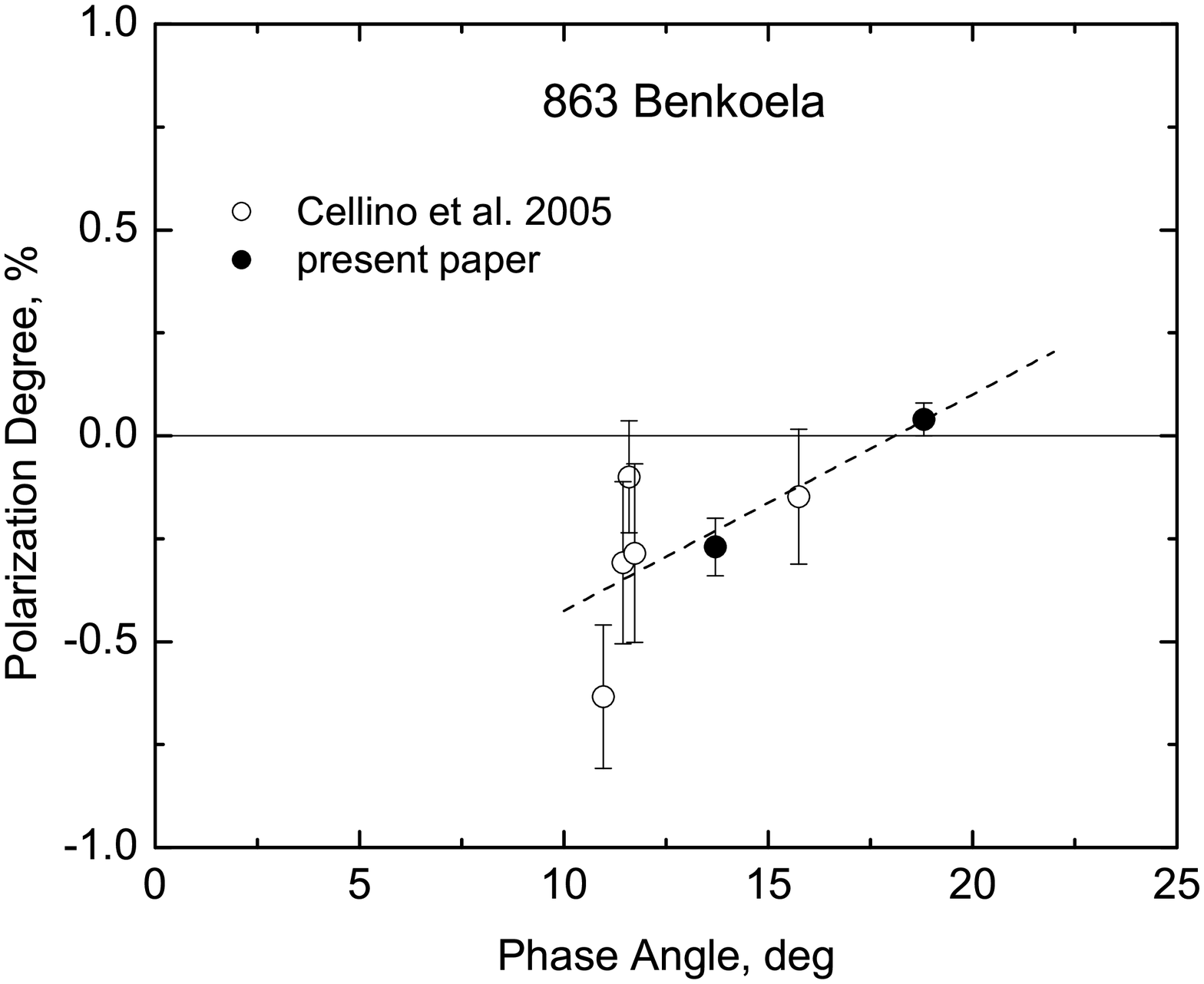}
   \caption{Polarization phase curve for the A-type asteroid 863 Benkoela.}
 \label{fig5}%
    \end{figure}

\subsection{Low albedo asteroids}

Asteroids of C- B-, F-, G-, and P-types have been previously observed. However, very few objects have been observed in each class, and thus it is difficult to distinguish
between their polarimetric properties, with the notable exception of the  F-type asteroids (see Fig. 6).
Recently, it was shown that observed F-type asteroids are characterized by substantially smaller depths of the negative polarization branch, as compared to other low albedo asteroids. In addition, they have unusually small inversion
angles (Belskaya et al. 2005).  Therefore, we have included several F-asteroids in our sample to check whether the
low inversion angles are inherent to all F-type asteroids. We find that 1021 Flammario has a smaller inversion angle than is typical for asteroids. At the same time, our observations of the near-Earth asteroid
3200 Phaethon, classified as F-type by Tholen (1985), definitely exclude a small inversion angle for it (see Fig. 6).  Note, however, that there is an uncertainty in the classification of this asteroid. Judging from its relatively high IRAS albedo (0.098), Phaethon probably belongs to the B class (Bus \& Binzel 2002).
Moreover, Licandro et al. (2005) found an absorption in the UV spectra of Phaethon inconsistent with the F
class. \\
Our polarimetric observations of other F-type asteroids, 225 Henrietta, 438 Zeuxo, 762 Pulcova, and 893 Leopoldina, do not give definite estimates of their inversion angles. We can only conclude that the
inversion angles of asteroids 438 Zeuxo and 762 Pulcova are larger than 16$^{o}$.

Among the other low albedo asteroids observed in our survey, an interesting case is represented by the P-type asteroid 420 Bertholda, which shows negative polarization of 1.03\% at the phase angle close to the angle of minimum polarization (Fig. 6). Such a polarization degree implies moderate surface albedo for this asteroid, in contradiction with its IRAS albedo of 0.042. An error in the albedo determination is unlikely since Bertholda belongs to the
Cybele group populated by low albedo asteroids. The probable explanation is that we observed a
decrease in the depth of the negative polarization branch for very dark surfaces, previously found for
the asteroid 419 Aurelia, which has an IRAS albedo of 0.046 (Belskaya et al. 2005).

We also made the first polarimetric observations of D-type asteroids, showing that the polarization value of the unusual minor planet 944 Hidalgo is similar to that of low albedo asteroids (Fig. 6).

\begin{figure}
   \centering
    \includegraphics[width=10cm]{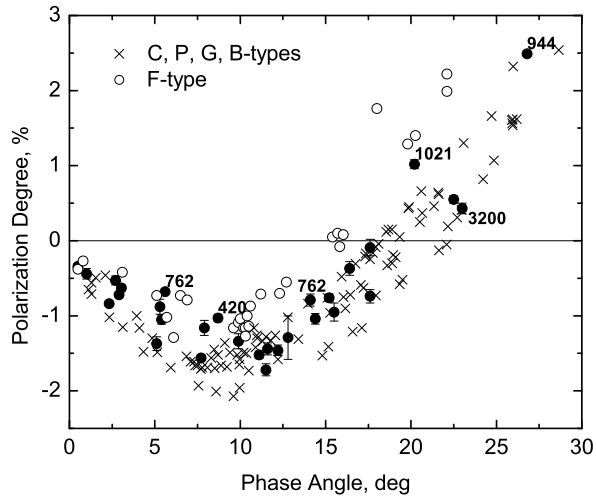}
   \caption{Polarization phase curve for low albedo asteroids. Data marked by crosses and open circles are taken from Zellner \& Gradie (1976), Belskaya et al. (2003, 2005), Chernova et al. (1994), Lupishko et al. (1994), and Cellino et al. (2005).}
 \label{fig6}%
    \end{figure}

\subsection{Small phase angle observations of 1021 Flammario}
\begin{figure}
   \centering
    \includegraphics[width=10cm]{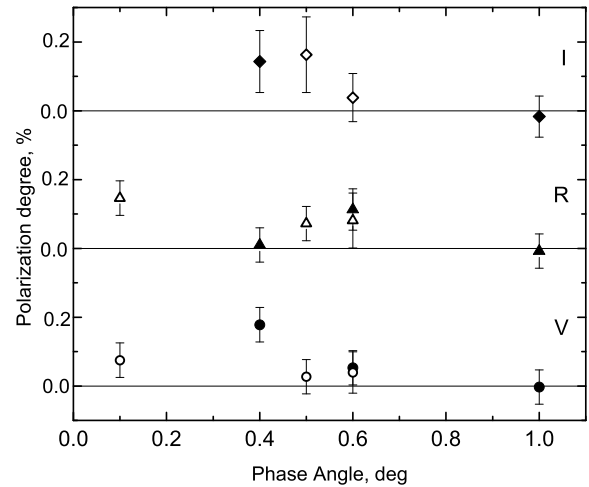}
   \caption{Polarization of 1021 Flammario at small phase angles. Open and filled symbols represent, respectively, observations before and after the asteroid's opposition.}
 \label{fig7}%
    \end{figure}

Detailed VRI observations of the low albedo asteroid Flammario have been carried out at extremely small
phase angles down to 0.1$^{o}$ (Table 2). These are the first measurements in such detail near opposition for a low albedo
asteroid. Measured values are shown in Fig. 7, where observations before and after opposition are marked by different symbols. The observations show zero or slightly positive polarization at small phase
angles in all filters, both before and after opposition. The observations at the same phase angle
of 0.6$^{o}$ made before and after opposition are in very good agreement.
The values of Q and U are close to each other at the same phase angles, indicating a $45^{o}$ deviation of the polarization plane from the scattering plane. This unusual characteristic could be attributed to an
instrumental effect, or else it could be related to physical properties of the asteroid surface.
Although the polarization degree is very small in the phase angle range, given the high accuracy of our calibration we exclude a systematic
error.  We favor an explanation related to the effect of surface anisotropy, e.g., an anisotropy caused by grooves, such us the Phobos grooves. Laboratory measurements
(Geake et al. 1984) have shown that this explanation is physically sound.

\section{Conclusions}

Using the AFOSC polarimeter, we have greatly expanded the number
of objects observed with high accuracy. Our survey includes near-Earth and main belt asteroids, as well as a
Jupiter-crossing object, for a total of 36 objects. The main
portion of the obtained polarimetric values is within the typical
range of average polarization-phase curves of a corresponding
taxonomic class. Our results prove that most asteroids have a
rather uniform surface microstructure. Only five asteroids in our
data set, namely 71 Niobe, 152 Atala, 420 Bertholda, 785 Zwetana,
and 1627 Ivar, are outside the typical values of their classes.
This result can derive either from an erroneous taxonomic
classification or from their particular surface properties.
Further data are needed to clarify the situation.

Small phase angle observations
of the asteroid 1021 Flammario show abnormal positive polarization in the range of phase angles 0.1-1.0$^{o}$, which can be attributed to surface structure anisotropy of the asteroid.

\begin{acknowledgements}
We thank Dr. A. Della Valle for his help in performing part of the observations
\end{acknowledgements}


\begin{thebibliography}{}
\bibitem[2005]{bar05} Barucci, M. A., Fulchignoni, M., Fornasier, S., et al. 2005, \aap, 430, 313--317
\bibitem[1985]{bel85} Belskaya, I.N., Efimov, Ju.S., Lupishko, D.F., \& Shakhovskoj, N.M. 1985, Sov. Astron. Lett., 11,  116--118
\bibitem[1987]{bel87} Belskaya, I.N., Kiselev, N.N., Lupishko, D.F., \& Chernova, G.P. 1987a, Kinemat. Phys. Neb. Tel, 3, 19--21
\bibitem[1987]{bel87b} Belskaya, I.N., Lupishko, D.F., \& Shakhovskoj, N.M. 1987b, Sov. Astron. Lett., 13, 219--220
\bibitem[1991]{bel91} Belskaya, I.N., Kiselev, N.N., Lupishko, D.F., \& Chernova, G.P. 1991, Kinemat. Phys. Neb. Tel, 7, 11--14
\bibitem[1996]{bel96} Belskaya, I.N., \& Lagerkvist, C. I. 1996, \planss, 44, 783
\bibitem[2003]{bel03} Belskaya, I.N.,  Shevchenko, V.G., Kiselev, N.N., et al. 2003, Icarus, 166, 276--284
\bibitem[2005]{belskaya05} Belskaya, I.N.,  Shkuratov Yu. G,  Efimov Yu. S., et al. 2005, Icarus, 178, 213--221
\bibitem[2004]{birlan04} Birlan M., Barucci M. A., Vernazza P., et al. 2004, New Astron., 9, 343--351
\bibitem[1990]{bro00} Broglia, P., \& Manara, A. 1990, \aap, 237, 256--258
\bibitem[1992]{bro92} Broglia, P., \& Manara, A., 1992, \aap, 257, 770--772
\bibitem[1994]{bro94} Broglia, P., Manara, A., \& Farinella, P. 1994,  Icarus, 109, 204--209
\bibitem[1985]{bus02} Bus, B., \& Binzel, R.P. 2002, Icarus, 158, 146-177
\bibitem[2005]{cellino05} Cellino, A., Hutton, R. G., di Martino, M., et al. 2005, Icarus, 179, 304--324
\bibitem[1999]{cellino99} Cellino, A., Hutton, R. G., Tedesco, E. F., di Martino, M., \& Brunini, A. 1999, Icarus, 138, 129-140
\bibitem[1994]{cher94} Chernova, G.P., Lupishko, D.F., \& Shevchenko, V.G. 1994, Kinemat. Phys. Neb. Tel, v. 10, No. 2, 45--49
\bibitem[2003]{delbo03} Delb\'o, M., Harris, A. W., Binzel, R. P., \& Pravec, P. 2003, Icarus, 166, 116--130
\bibitem[2004]{desidera04} Desidera, S., Giro, E., Munari, U., et al. 2004, \aap, 414, 591--600
\bibitem[2006]{fornasier06} Fornasier, S.,  Belskaya, I., Fulchignoni, M., Barucci, M. A., Barbieri, C. 2006, \aap, 449, 9--12
\bibitem[1984]{geake04} Geake, J., Geake, M., \& Zellner, B. 1984, \mnras, 210. 89-112
\bibitem[1974]{gehrels74} Gehrels, T., 1974, in {\it Planets, stars and nebulae studied with photopolarimetry},  ed. T. Gehrels (Tucson: University of Arizona Press), 3
\bibitem[1994]{kiselev94 } Kiselev, N.N., Chernova, G.P., \& Lupishko, D.F. 1994,  Kinematika. Fiz. Nebesnyka Tel., 10, 35--39
\bibitem[2005]{licandro05} Licandro, J., 2005, personal communication
\bibitem[1994]{lupishko94} Lupishko, D. F., Kiselev, N.N., Chernova, G.P., Shakhovskoj, N.M., \& Vasilyev, S.V. 1994, Kinemat. Phys. Neb. Tel, v. 10, No. 2,
40--44
\bibitem[1996]{lupishko96} Lupishko, D. F. \&  Mohamed, R. A. 1996, Icarus, 119, 209--213
\bibitem[1999]{mag} Magri, C., Ostro, S.J., Rosema, D.K., et al. 1999, Icarus, 140, 379--407
\bibitem[2002]{mui02} Muinonen, K., Piironen, J., Shkuratov, Yu. G., Ovcharenko, A., \& Clark, B. E. 2002, in {\it Asteroids III}, ed. W. Bottke, R.P. Binzel, A. Cellino, \& P. Paolicchi (Tucson: University of Arizona Press), 1
\bibitem[2004]{mui04} Muinonen, K., 2004, Waves in Random Media, 14, Iss. 3, 365
\bibitem[2005]{pentilla05} Pentilla A., K. Lumme, E. Hadamcik, \& Levasseur-Regourd, A.C. 2005, \aap, 432, 1081--1090
\bibitem[2003]{pernechele03} Pernechele C., Fantinel D., \& Giro E. 2003, SPIE Proc., 4843, 156--163
\bibitem[2000]{riv}Rivkin, A.S., Howell, E.S., Lebofsky, L.A., et al. 2000, Icarus, 145, 351--368
\bibitem[1972]{sh72} Shakhovskoy N. M. \& Efimov, Yu. S. 1972, Izv. Krymskoi Astrofiz. Obs., 45, 90--110
\bibitem[1974]{serkowski74} Serkowski, K. 1974, in {\it Planets, stars and nebulae studied with photopolarimetry},  ed. T. Gehrels (Tucson: University of Arizona Press), 135
\bibitem[2002]{sh02}Shkuratov, Yu.G., Ovcharenko, A., Zubko, E., et al 2002, Icarus, 159, 396--416
\bibitem[2004]{sh04}Shkuratov, Yu., Ovcharenko, A., Zubko, E., et al. 2004, J. Quant. Spectr. Rad. Trans., 88, 267--284
\bibitem[2002]{tede02} Tedesco, E. F., Noah, P. V., Noah, M., \& Price, S. D. 2002, \aj, 123, 1056--1085
\bibitem[1985]{tholen85} Tholen, D. J. 1985, IAU Circular No. 4034
\bibitem[1989]{tholen89} Tholen, D. J. 1989, in {\it Asteroids III}, ed. W. Bottke, R.P. Binzel, A. Cellino, \& P. Paolicchi (Tucson: University of Arizona Press), 1139
\bibitem[2001]{volten01} Volten, H., Munoz, O., Rol, E., et al. 2001, J. of Geoph. Res., 106, Iss. 17, 375--401
\bibitem[1976]{zellner} Zellner, B., \&  Gradie J. 1976, \aj, 81, 262--280
\bibitem[2005]{zub05} Zubko, E., Petrov, D., Shkuratov, Y., \& Videen, G. 2005, Applied Optics, 44, 6479--6485


\end{thebibliography}
\end{document}